\documentclass[english,aps,prl,twocolumn,showpacs,groupaddress,floatfix,graphics,graphicx]{revtex4-1}
\usepackage[latin9]{inputenc}
\usepackage{amsmath}
\usepackage{amssymb}
\usepackage{graphicx}
\usepackage{verbatim}
\usepackage{braket}
\usepackage{soul}
\usepackage{gensymb}
\usepackage[usenames]{color}

\renewcommand\vec{\boldsymbol}

\begin{document}

\title{Current Driven Magnetization Reversal in Orbital Chern Insulators} 

\author{Chunli Huang, Nemin Wei, and Allan~H.~MacDonald}
\affiliation{Department of Physics, University of Texas at Austin, Austin TX 78712}

\date{\today} 

\begin{abstract}
Graphene multilayers with flat moir\'e minibands can exhibit the quantized anomalous Hall 
effect due to the combined influence of spontaneous 
valley polarization and topologically non-trival valley-projected bands.
The sign of the Hall effect in these Chern insulators can be reversed either by applying an external 
magnetic field, or by driving a transport current through the system.  
We propose a current-driven mechanism whereby reversal occurs along lines in 
the (current $I$, magnetic-field $B$)
control parameter space with slope $dI/dB = (e/h)\, M A_{M} \, (1-\gamma^2)/\gamma$, 
where $M$ is the magnetization, $A_M$ is the moir\'e unit cell area, and $\gamma <1$ is the ratio of the 
chemical potential difference between valleys along a domain wall to the electrical bias $eV$. 
\end{abstract}

\maketitle

{\it Introduction:}---
Magnetism in solid state system is produced by both spin and orbital electronic angular momentum, but the 
two constituents normally have a decidedly asymmetric relationship in which spins order spontaneously
and orbital magnetism is induced parasitically by spin-orbit interactions.  
Current control of ordered spins is now routine in spintronics \cite{bader2010spintronics,bhatti2017spintronics,kruglyak2010magnonics,roche2015graphene}.  
The recent discovery \cite{sharpe2019emergent, serlin2020intrinsic} of spontaneous orbital order 
manifested by a quantum anomalous Hall effect in graphene moir\'e superlattice systems, 
and of current driven magnetization reversal in those systems, is the first demonstration of,
an influence of a transport current on orbital magnetism.
In this Letter we propose an experimentally testable explanation for this effect.

The quantum anomalous Hall effect, a property of insulators whose occupied bands carry a net 
Chern number, is common 
in graphene moir\'e superlattice systems   \cite{sharpe2019emergent, serlin2020intrinsic,chen2019tunable,polshyn2020nonvolatile,tschirhart2020imaging} when the minibands are flat and the moir\'e
band filling factor  $\nu = n_e A_M$ is close to an odd integer.
(Here $n_e$ is the carrier density and $A_M$ is the moir\'e unit cell area.)
In magic angle twisted bilayer graphene \cite{bistritzer2011moire} (MATBG), for example,
the intriguing family of strongly correlated states in the $ -4 < \nu < 4 $ flat-band regime includes 
superconductors and Mott insulators\cite{cao2018correlated,cao2018unconventional,lu2019superconductors,yankowitz2019tuning},
and also a Chern insulator state with a Hall resistance close \cite{sharpe2019emergent, serlin2020intrinsic}
to the von Klitzing constant.  The quantized Hall conductance appears at $\nu=3$ when the graphene bilayer
is aligned with an adjacent hexagonal boron nitride layer, but
unlike the case of magnetized topological insulators \cite{nagaosa2010anomalous,chang2013experimental,he2018topological},
cannot be a consequence of spin-order plus spin-orbit coupling 
since the latter is negligible in pristine graphene.
The Chern insulator is instead thought to be 
a combined consequence of the non-trivial topology of moir\'e minibands in graphene multilayers \cite{bultinck2020mechanism,wu2020collective,liu2020correlated,kim2017tunable,alavirad2019ferromagnetism,repellin2020ferromagnetism,zhang2019twisted,xie2020nature,zhang2019nearly} and
momentum-space condensation \cite{Heisenberg1984,london1948problem,jung2011lattice} in
the form of spontaneous valley polarization.  
Indeed, Hartree-Fock calculations \cite{xie2020nature,hejazi2020hybrid} predict that
odd integer $\nu$ insulators in graphene multilayers
are very often Chern insulators.  We refer to these states as orbital Chern insulators (OCIs) although
they break time reversal symmetry in both spin and orbital degrees of freedom,
because the main observable - the anomalous Hall effect - is of orbital 
origin, and because spin-order cannot be maintained at 
finite temperature when spin-orbit interactions are negligible.
We therefore drop the spin-degree of freedom from the following discussion.
The properties of OCIs are quite distinct\cite{jihang2020} from those of 
spin Chern insulators \cite{he2018topological}.
From a statistical physics point of view, an OCI is an Ising ferromagnet in which 
the total Chern number of the occupied bands $C_{\pm}=\pm C$ can be viewed as an order parameter. 
 
\begin{figure*}
    \centering
   \includegraphics[width=2.0\columnwidth]{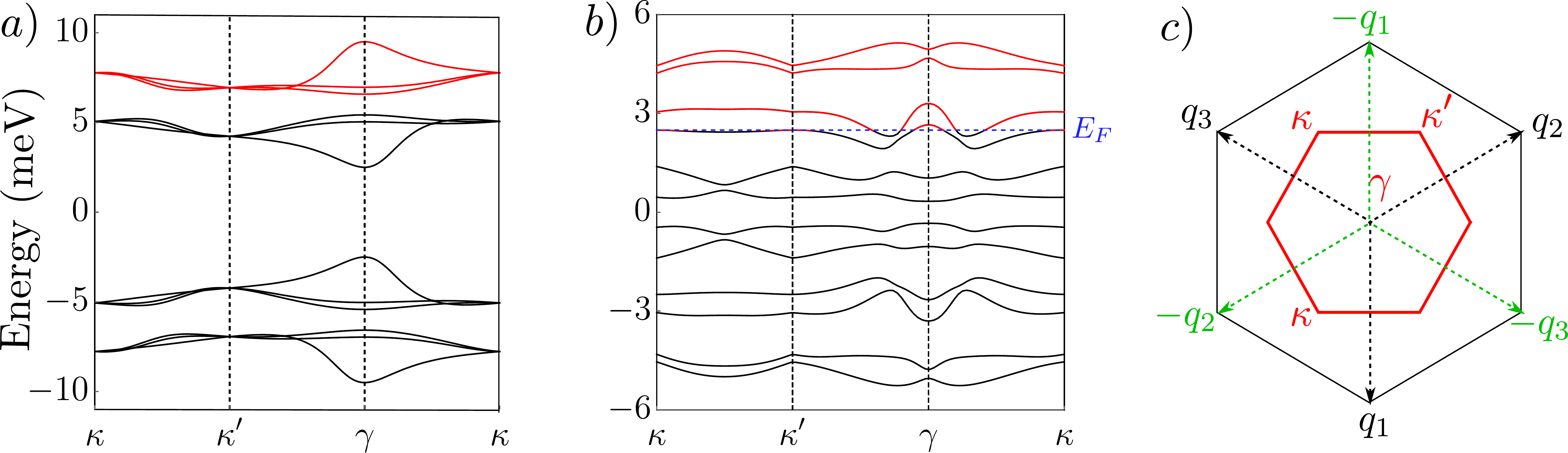}
    \caption{
Local quasiparticle bands (Eq.~\eqref{eq:H_MF}) for a 
valley-exchange field pointing along a) $\hat{n}=\hat{z}$ and b) $\hat{n}=\hat{x}$. 
Occupied (unoccupied) states at $\nu=n_e A_M=3$ are drawn in black (red) so that a) is an insulator 
while b) is a metal.
For better comparison, the bands in a) have been folded into the irreducible Brillouin zone of b). 
Because the inter-layer tunneling terms are different in the two valleys (c) the
area of the irreducible Brillouin zone (red) for general valley orientation
(spanned by $\vec{q}_1$ and $\vec{q}_2$) is 1/3  of the single-particle valley-projected 
moir\'e Brillouin zones (black-solid) area (spanned by $\vec{q}_2-\vec{q}_1$ and $\vec{q}_3-\vec{q}_1$).
These bands were calculated using spontaneous valley splitting 
$I=12$meV, hBN-induced mass gap $\Delta_{\text{BN}}=10$meV, 
twist angle $\theta=1.05^{\circ}$, and Fermi velocity $v_{F}=9.5\times 10^{5}\text{m/s}$.}
    \label{fig:band}
\end{figure*}

Experiments have shown that the Hall conductance of an OCI can be switched between $+Ce^2/h$ and $-Ce^2/h$,
signaling a complete reversal of orbital magnetization \cite{sharpe2019emergent,  serlin2020intrinsic},
by applying either an external magnetic field $B$ and/or an electrical bias voltage $V$. 
The magnetization reversal mechanisms in conventional spin ferromagnets  
are relatively well established  \cite{malozemoff2016magnetic,hubert2008magnetic,upadhyaya2016domain}, and involve a combination of Stoner-Wohlfarth single domain switching
and domain-wall depinning, driven by a combination of spin-transfer torques, spin-orbit torques, and magnetic fields.
Consensus has however not yet been reached on the microscopics of orbital-magnetization reversal,
although some interesting proposals have been put forward \cite{su2020,he2020giant,serlin2020intrinsic}.
Here we analyze the case of current driven reversal in an OCI with a bulk that 
is perfectly insulating so that gapless charge excitations are present only at the 
sample edge and along domain walls.  
We find that both magnetic fields $B$ and transport bias voltages $V$
apply pressure to domain walls and predict that switching occurs along a line in
the (current $I$, magnetic-field $B$)
control parameter space with slope $dI/dB = (e/h)\, M A_{M} \, (1-\gamma^2)/\gamma$, where $M$ is the 
magnetization, $A_M$ is the moir\'e unit cell area, and $\gamma <1$ is the ratio of the 
chemical potential difference between valleys along a domain wall to the electrical bias voltage. 
In the following we first argue that moir\'e superlattice OCIs are described by an $O(3)$ field 
theory in which the vector order parameter characterizes the local valley polarization
direction.  This property allows domain pinning 
to be analyzed using conventional Landau-Lifshitz equations.

{\it Valley-pseudospins in MATBG:}---
 The valley-projected
$\pi$-bands of twisted bilayer graphene are described by a low-energy continuum model \cite{bistritzer2011moire} 
in which isolated layer Dirac cones are coupled by an inter-layer tunneling term that 
has the periodicity of the moir\'e pattern: 
\begin{align} \label{eq:H_BM}
H_0^{\tau} =& -i\hbar v_F(\tau \sigma_x\partial_x + \sigma_y \partial_y) -\frac{\Delta_{\text{BN}}}{2} \sigma_z |1\rangle \langle 1| \nonumber \\ &+ \sum_{j=1}^{3} \big[ T_{j}^{(\tau)}e^{-i\tau \vec{q}_{j}\cdot\vec{r}}
|1\rangle \langle 2| + h.c.\big],  \\ 
T_j^{\tau}=&w_0 \sigma_0 + w_1\sigma_x e^{\frac{2\pi i}{3}(j-1)\tau \sigma_z},
\label{eq:band}
\end{align}
where $\tau =\pm$ is the valley label, $|1\rangle \langle 2|$ accounts for tunneling 
between layers labelled $1$ and $2$, the $\vec{\sigma}$ Pauli matrices
act on the sublattice degree of freedom within each layer, 
$q_1=(4\pi/3a_M)(0,-1)$ , $q_{2,3}= (4\pi/3a_M) (\pm\sqrt{3}/{2},1/2)$ 
and $a_M$ is the moir\'e lattice constant, equal to $13.4$nm at the magic angle $\theta=1.05^{\circ}$.  
In Eq.~\ref{eq:H_BM}
$w_0,w_1$ are tunneling energy parameters whose values are known.  
Since $H_0^{\tau}$ is a periodic function of position for each valley $\tau$, 
it has a set of Bloch bands 
$
 H_0^{\tau}  |u_{n\tau\vec{k}}\rangle
= E_{n\tau}(\vec{k}) |u_{n\tau\vec{k}}\rangle
$
that satisfy the time-reversal symmetry property $E_{n-}(-\vec{k}) = E_{n+}(\vec{k})$,
guaranteeing that the densities of states of the two-valleys are identical. 

The OCI ground state at $\nu=3$ empties the conduction band of one 
valley, chosen spontaneously.  Mean-field calculations \cite{xie2020nature} have shown that the energy 
scale $I$ of single-particle valley-flip excitations of the OCI 
state is $\sim 10$meV, whereas the energy scale $K$
of long-wavelength collective valley reorientation excitations \cite{ajesh2020}$\sim 0.1$ meV.
This contrast in energy scales is familiar from the properties of
the conventional itinerant electron ferromagnets 
heavily employed in spintronics, although less extreme in the OCI case, if we identify valley in OCIs with spin in conventional 
ferromagnets. (In ferromagnetic Ni for example $I \sim 0.3$eV and $K \sim 3 \mu$eV
\cite{fritsche1987relativistic}.)  We therefore follow the approach used in metal spintronics to 
address magnetization reversal by assuming that we can focus on the dynamics of the low-energy 
collective degrees of freedom, which are described at long wavelengths by the phenomenological 
micromagnetic \cite{micromagnetics} energy density:
\begin{align} \label{eq:sigma_model}
\mathcal{E}[\vec{n}]= A (\nabla \vec{n})^2- KA_M n_z^2 + K_{\perp}A_M \sin^2(\theta) \sin^2(\phi-\phi_p)
\end{align}
where $\hat{n}=(\sin\theta \cos \phi,\sin \theta \sin \phi,\cos \theta )$ is the Bloch sphere unit vector that  
characterizes the local collective valley spinor
\begin{equation}
\label{eq:mixingphase}
|\Psi\rangle\sim
\cos\left(\frac{\theta}{2}\right)\, |\tau=+\rangle \,+\, e^{i\phi} \sin\left(\frac{\theta}{2}\right)  \,|\tau= -\rangle.
\end{equation}
Eq.~\eqref{eq:sigma_model} is parameterized by three parameters (with dimenson of energy) $A,K,K_{\perp} >0$ which arise naturally from the following considerations:
$A$ is a stiffness parameter that expresses an energetic preference for uniform valley polarized states, 
$K$ is a valley anisotropy constant that favors complete polarization in $|\tau=\pm\rangle$  in the OCI ground state,
and $K_{\perp}$ is an azimuthal anisotropy constant that accounts for processes that violate 
valley conservation. Since $K_{\perp}=0$ is a consequence of momentum conservation in perfect crystals, we anticipate that
$K_{\perp} \neq 0$ only near sample edges.

In Fig.~\ref{fig:band} we plot the mean-field quasiparticle energy 
bands of an OCI for two different valley-orientations $\hat{n}$ by adding an exchange field
with Stoner interaction constant $I=12$ meV to Eq.~\eqref{eq:band}
\begin{equation}
H_{MF} = \frac{H_0^{+}+H_0^{-}}{2} + \tau_z \;  \frac{H_0^{+}-H_0^{-}}{2} - \frac{I}{2} \vec{\tau} \cdot \vec{n}  \label{eq:H_MF},
\end{equation}
where the $\vec{\tau}$ Pauli matrices act on the valley degree of freedom.  The choice of 
an exchange effective magnetic field that is aligned with the valley orientation is motivated 
by the observation the dominant Coulomb interactions in graphene multilayers are valley-independent, 
just as the Coulomb interactions in a magnetic metal are spin-independent.  
The OCI band-structure calculation has three important messages. 
First, the bandstructure is independent of $\phi$ as a result of total valley number conservation
in Eq.~\ref{eq:H_MF}.  This band model result is consistent with the expectation that 
$K_{\perp}=0$ in perfect periodic lattice. Second, as $\hat{n}$ goes from the pole (Fig.~1a) to the equator (Fig.~1b), the bandwidth decreases and total energy increases, suggesting that the easy direction of valley
polarization is the polar axis in agreement with experiment. 
Because the exchange field couples valleys and the tunneling Hamiltonians in the two
valleys are not identical, the moir\'e Hamiltonian unit cell area is increased 
(by a factor of three) when $\sin(\theta/2) \ne 0$, as illustrated in 
Fig.~1c. Third, we found that the local band structure 
is metallic when $\hat{n}$ is close to the equator, a property that will have
important implications for domain wall dynamics.  


{\it Domain Wall Dynamics:--}
We can calculate magnetization dynamics from Eq.~\ref{eq:sigma_model} by recognizing that
the two-components of valley pseudospin perpendicular to $\hat{n}$ (when suitably normalized) are canonical
conjugate variable.  The Euler-Lagrange equation corresponding to Eq.~\ref{eq:sigma_model}
is therefore:
\begin{equation} \label{eq:LLG}
\frac{\hbar }{2 A_M} \partial_t \vec{n}= 
{\vec n}\times\frac{\delta \mathcal{E}}{\delta{\vec n}}+ \frac{\hbar }{2 A_M}  \,{\vec n}\times ( \check{\alpha} \cdot \partial_t {\vec n}).
\end{equation}
Eq.~\ref{eq:LLG} is known in spintronics as the Landau-Liftshitz Gilbert equation and 
includes a damping tensor $\check{\alpha}$ that accounts for coupling between collective magnetic degree of freedom
and other low-energy degrees of freedom, including phonons and gapless quasiparticle excitations if these are present.
Applying Eq.~\ref{eq:LLG} to Eq.~\ref{eq:sigma_model} and linearizing around $n_z=1$ yields the 
valley-wave collective mode energies
$
E(q) = 4K +  4 A A_M q^2 .
$
By fitting to microscopic bulk collective mode calculations \cite{ajesh2020}, we 
estimate that $K \sim 0.04$meV and $A A_M \sim 0.13\text{meV}a_{M}^2$. 

\begin{figure}[t]
\includegraphics[width=1\columnwidth]{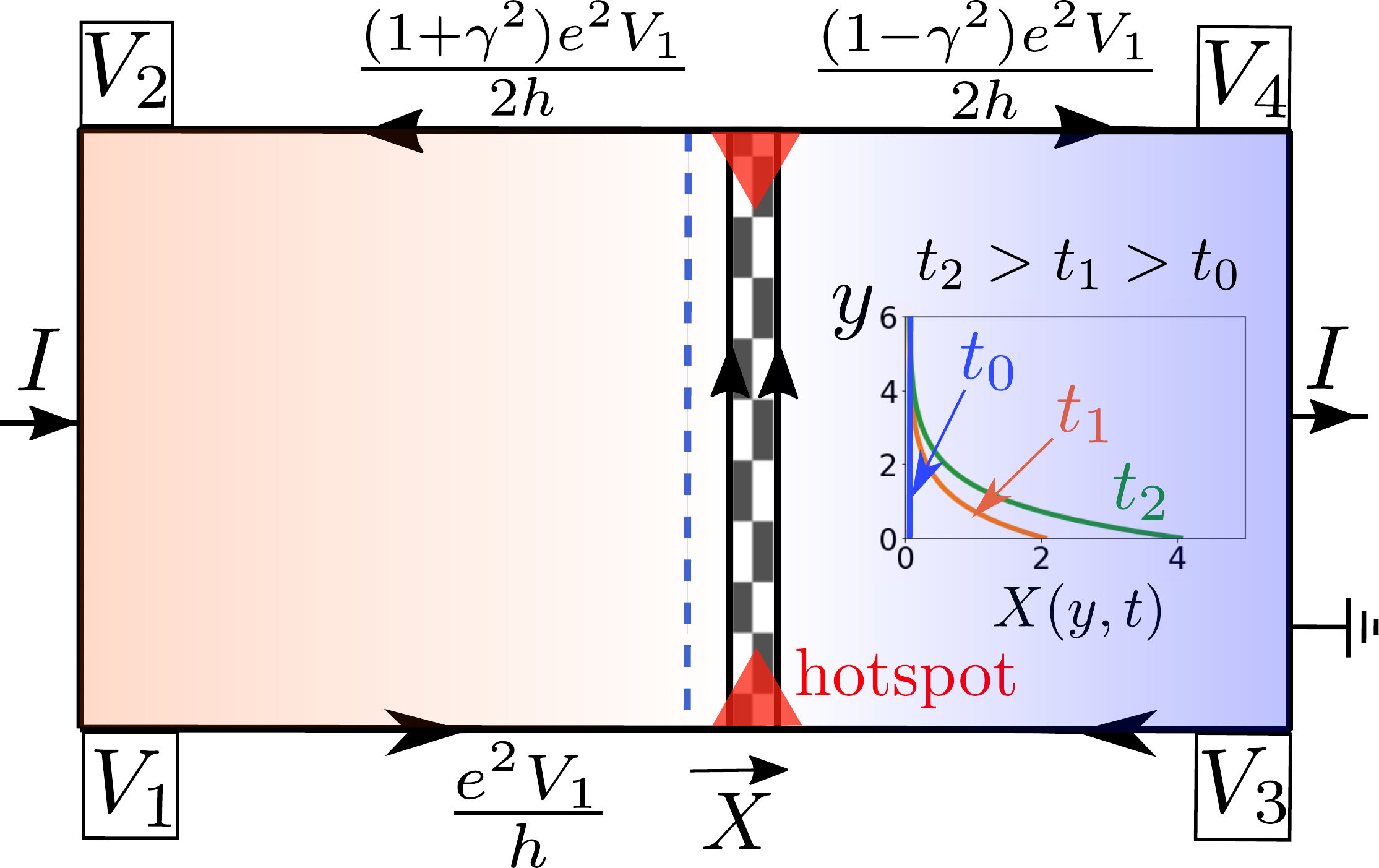}
\caption{Magnetization reversal in an orbital Chern insulator: 
A domain wall (vertical dashed line) 
separates the OCI into regions with
opposite signs of the Hall conductances.  Domain walls can be shifted (from $0$ to $X$) and eventually depinned by a valley-dependent 
chemical potential $e(V_1-V_3)$, or by a magnetic field. Inset shows bending of a domain wall close to a hotspot \cite{SM}.} \label{fig:drawing}
\end{figure}

A domain wall, like the one illustrated schematically in Fig.~\ref{fig:drawing},
is a real-space topological defect obtained by minimizing the energy 
functional Eq.~\ref{eq:sigma_model} with the constraint that $n_z \to \pm 1$ for $x \to \pm \infty$.
This yields $\phi = \phi_p$ and $  \theta=2\arctan\,\text{exp}\left( x-X/\lambda \right),$
where $X$ is the domain wall center, $\lambda = \sqrt{AA_M/K}$ is half-width of the domain wall.
Using the values for $K$ and $A$ quoted above yields $\lambda=1.8a_{M}$.  
In order to describe wall dynamics, we use a generalization of
Slonczekswi's \cite{slonczewski1972dynamics} \textit{ansatz} by letting the domain wall
position and azimuthal phase,
\begin{equation}
X=X(y,t) \; ,\; \phi=\phi(y,t),
\end{equation}
depend on time and the coordinate along the wall.
This dynamics focuses on excitation of the {\it soft-mode} of a domain wall
associated with its invariance under a shift in $X$ in the absence of pinning.  
In practice, domain walls are invariably pinned by sample 
inhomogeneities in real devices, and this pinning is responsible for hysterisis.
For definiteness we assume that 
the domain wall is pinned at $X=0$ by some extrinsic pinning potential $E_{pin}$ which can arise from, 
{\it e.g.~} a twist-angle extremum at which the condensation energy of the ordered state 
is minimized, or a local minimum in the width $W$ of the sample. 
There is an energy penalty $dE_{pin}$ to shift $X$ away from $X=0$.  
For simplicity, we take it to be specified by a harmonic potential
\begin{equation} \label{eq:Epin}
dE_{pin} = E_{pin}(X)-E_{pin}(X=0)=\frac{kW}{2}X^2,
\end{equation}
up to a maximum $|X|<X_{max}$ beyond which the pinning energy is constant. 
The pinning strength $k>0$ has units of energy per length. Our main results do
not depend on the details of $E_{pin}$.  When an external magnetic field $B$ is present 
we must also account for the dependence of its interaction with the spontaneous orbital
magnetization on domain wall position,
\begin{equation} \label{eq:EB}
dW^{B}_{\pm} = [- M_{+}X -M_{-}(-X)]WdB  = - 2 M WX \,dB.
\end{equation}
Here $M_{\pm}$ is the net orbital moment per area of the $\pm$ valley states 
and we used time reversal symmetry ($M_{+}=-M_{-}\equiv M$) in the second equation. 
Introducing Eqs.~\eqref{eq:Epin} and \eqref{eq:EB} into Eq.~\eqref{eq:sigma_model} and integrating 
Eq.~\eqref{eq:LLG} over $x$ yields: 
 \begin{align} \label{eq:X_n_phi}
\dot{\phi}  &= \frac{4AA_M}{\hbar \lambda} X'' +
\frac{2 A_M M B}{\hbar} -\frac{A_M k X}{\hbar}-  \alpha_{\phi}\frac{\dot{X}}{\lambda}, \\
 \frac{ \dot{X}}{ \lambda}	&= -\frac{2AA_{M}\pi}{\hbar}\phi'' + \frac{ 2 K_{\perp}}{\hbar}\sin(2(\phi-\phi_p)) + \alpha_X \dot{ \phi} .
 \label{eq:X_n_phi_2}
 \end{align}
Eq.~\eqref{eq:X_n_phi} equates the precession frequency of the valley pseudospin to the wall-pressure 
generated by the sum of wall-curvature, magnetic-field, pinning forces, and damping forces.
Note we distinguished $\alpha_X$ from $\alpha_\phi$
since it requires processes that change overall valley polarization, and we therefore expect it to be much smaller. Indeed, $\alpha_\phi$ has a substantial electronic contribution
since \cite{kambersky1976ferromagnetic} the 
Chern number change upon valley polarization reversal requires that the quasiparticle gap vanishes in the interior of the domain wall, see Fig.~1b).
Eq.~\eqref{eq:X_n_phi_2} is a continuity equation (with $K_{\perp}\rightarrow 0$) for valley-polarization
expressed in collective coordinates: the damping term proportional to $\alpha_X$ is a valley-transfer torque that accounts for the valley-pumping 
quasiparticle currents generated by $\dot{\phi}$ \cite{tserkovnyak2005nonlocal,tserkovnyak2002enhanced,tsoi1998excitation,ralph2008spin,slonczewski1996current,berger1996emission,slonczewski1999excitation}.

So far we have not directly invoked the unusual physics of OCIs, except by 
allowing the valley polarization order parameter, which is important for identifying conjugate 
coordinates and therefore collective coordinate dynamics, and magnetization, 
which characterizes the strength of interactions with the external magnetic field, to be independent.
For spin-magnets these two 
quantities have a universal relationship characterized by the gyromagnetic ratio.  
The simple way in which 
transport currents influence domain wall dynamics, which we now explain, is however a very specific consequence of 
the topological character of OCIs.

The pinned domain wall in Fig.~\ref{fig:drawing} separates orbital Chern 
insulator domains with opposite total Chern numbers. The domain wall therefore supports 
two co-propagating edge channels that are sourced entirely from different 
electrical contacts when tunneling between channels is negligible.
We identify the local chemical potential difference between valleys on the domain wall 
with $\hbar \dot{\phi}$ via the 
Josephson-like voltage-frequency relationship \cite{berger1986possible}:  
\begin{equation} \label{eq:mag_jose}
 \hbar \dot{\phi}= \delta \mu
\end{equation}
This fundamental relationship allows topological edge states to electrically control the properties of OCIs.
Anticipating that substantial equilibration occurs in the hot-spot regions \cite{wei2018electrical,zhou2020solids,takei2016spinsuperfluidity}
indicated in Fig.~2 where 
the valley edge state channels meet near the sample boundary so that momentum is not-conserved, we 
set $\delta \mu = \gamma eV$, where $\gamma<1$ is a fractional equilibration 
parameter.

\textit{Quasistatic-wall:--}
Eq.~\ref{eq:X_n_phi} has a quasistatic solution with 
$\phi$ and $X$ independent of $y$, and $X$ independent of time:
\begin{equation} \label{eq:X_eq}
X_{\text{eq}} = \frac{2M B-A_M^{-1} \gamma eV}{k}.
\end{equation}
Eq.~\eqref{eq:X_eq} has the following thermodynamic interpretation.
The chemical potential is the energy to add an electron to the system.  
In an ordinary insulator chemical potentials within the gap are undefined because the
system is incompressible; no states are available to add electrons within the gap.
In a Chern insulator, electrons can be added at energies within the gap, but only at an edge
or a domain wall, and only by expanding the area of the system so that it holds one more electron. 
When a domain wall
moves it adds electrons to one Chern insulator and removes it from the other.  
Eq.~\ref{eq:X_eq} places the domain wall at the position where 
the the energy change for moving a domain wall 
by $A_M/W$, adding an electron to one domain and removing it from the other,
is the chemical potential difference
$\hbar  \dot{\phi}$.

Reversal occurs at the depinning threshold $X_{\text{eq}}=X_{max}$.
According to Eq.~\ref{eq:X_eq}, the slope of the $X_{\text{eq}}=X_{max}$ line in the $(\delta \mu,B)$ parameter space is 
\begin{equation}\label{eq:main}
\frac{d\delta\mu}{dB}= 2M A_M.
\end{equation}
To relate reversal to the transport current we note that
since two hot spots have been traversed the difference in local chemical potentials 
between top left and top right of the Hall bar in Fig.~\ref{fig:drawing} is 
$\gamma^2 eV$.  It follows that the net current flowing from 
source to drain is 
\begin{equation} 
I = V \frac{e^2}{h} \frac{1-\gamma^2}{2}
\end{equation}
and that reversal therefore occurs along a line in control parameter space with slope 
\begin{equation}\label{eq:main}
\frac{dI}{dB}= M A_M \frac{e}{h} \frac{1-\gamma^2}{\gamma}.
\end{equation}
This is the central result of the paper.
Since $MA_M \sim \mu_B$ in graphene multilayer OCIs \cite{polshyn2020nonvolatile} and $e \mu_{B}/\hbar \sim 1.4 \times 10^{-8}$A/T is two-orders of magnitude smaller than
the experimental result reported in Ref.~\onlinecite{serlin2020intrinsic},
this mechanism can apply to the samples studied experimentally only if $\gamma \sim 10^{-2}$, 
\textit{i.e.} the edge channels are substantially equilibrated in the
hot spot regions.  This property is in fact consistent with reported observations~\cite{serlin2020intrinsic}. Our theory of reversal can be tested quantitatively by
measuring\cite{yasuda2017quantized,rosen2017chiral} the longitudinal resistance
along the upper edge of the Hall bar in 
Fig.~\ref{fig:drawing} to determine a value for $\gamma$:
\begin{equation} 
R = \frac{V_2-V_4}{I} = \frac{2h}{e^2} \frac{\gamma^2}{1-\gamma^2}.
\end{equation}

\textit{Discussions:--}
When edge states arrive at a hotspot (cf.~Fig.~2) with a valley (momentum) flux perpendicular to the domain wall, they can exert a force on the domain wall.
If so, the wall will bend with a vertical profile satisfying Eq.~\eqref{eq:X_n_phi}--\eqref{eq:X_n_phi_2}, see Ref.~\onlinecite{SM}.
%
Illustrative wall profiles are shown in Fig.~2. Observation of such domain wall bending would support our proposed reversal mechanism.


Two interesting mechanisms for current reversal of orbital magnetization have 
recently been proposed in Refs.~\onlinecite{su2020} and 
\onlinecite{he2020giant}. Their theories appeal to finite dissipation in the bulk ($\sigma_{xx} \neq 0$) 
and do not apply in the quantum anomalous Hall effect regime considered here.
The theoretical analysis in Ref.~\cite{serlin2020intrinsic} identifies 
an $I^3$ contribution (where $I$ is the current) to the edge state free energy
of conduction edge states and associated reversal with it 
becoming comparable to bulk magnetostatic energy. 
This reversal mechanism does not rely on wall dynamics. 
Our theory provides an alternative current-reversal mechanism based on depinning of valley domain walls
via topological edge states and it is most relevant in devices with well defined orbital Chern insulators,
like those imaged recently in Ref.~\cite{tschirhart2020imaging}.



\textit{
Acknowledgement:}
The authors acknowledge informative interactions with Eli Fox, David Goldhaber-Gordon, Gregory Polshyn, 
Aaron Sharpe, and Andrea Young.  This work was supported by  DOE grant DE- FG02-02ER45958 and 
Welch Foundation grant TBF1473.

\bibliographystyle{ieeetr}
\bibliography{reference}

\pagebreak
\widetext
\begin{center}
\textbf{\large Supplementary Materials: Current Driven Magnetization Reversal in Orbital Chern Insulators}
\end{center}
\setcounter{equation}{0}
\setcounter{figure}{0}
\setcounter{table}{0}
\setcounter{page}{1}
\makeatletter
\renewcommand{\theequation}{S\arabic{equation}}
\renewcommand{\thefigure}{S\arabic{figure}}
\renewcommand{\bibnumfmt}[1]{[S#1]}
\renewcommand{\citenumfont}[1]{S#1}

In this supplementary material, we discuss the dynamics of orbital Chern insulator domain wall when an electric current pass through the system. As discussed in the maintext, the equation of motion for a domain wall located inside the pinning region is given by the following:
 \begin{align} \label{eq:SX_n_phi}
\dot{\phi}(y,t)  &= \frac{4AA_M}{\hbar \lambda} X''(y,t)  +
\frac{2 A_M M B}{\hbar} - \frac{A_M k }{\hbar} X(y,t) -  \alpha_{\phi}\frac{\dot{X}(y,t)}{\lambda}, \\
 \frac{ \dot{X}(y,t)}{ \lambda}	&= -\frac{2AA_{M}\pi}{\hbar}\phi''(y,t)  + \frac{2K_{\perp}}{\hbar} \sin2(\phi(y,t)-\phi_p)+\alpha_X \dot{ \phi}(y,t) .
 \label{eq:SX_n_phi_2}
 \end{align}
The equations above can be viewed as a dynamical system describing the time evolution of vector variable $\vec{\chi} = (\phi,X/\lambda)^{\text{T}}$. In the absence of electric current, domain wall is static and the minimum wall energy configuration is described by the following fixed points:
\begin{equation}
  \left( \phi^*, \frac{X^*}{\lambda} \right)^{\text{T}} = \left( \phi_p+ \frac{n\pi}{2} , \frac{2MB}{k\lambda} \right)^{\text{T}} \equiv  \vec{\chi}^* 
\end{equation}
where $n \in \mathbb{Z}$. 

When an electric current pass through an orbital Chern insulator with a domain wall (c.f.~Fig.~2), it has to traverse along the domain wall by populating topologically protected edge states. We assume the current-carrying states arriving at the hotspot from contacts will partially equilibrate and enter the domain wall with a finite valley chemical potential $\delta \mu$. If so, $\delta \mu$ will precess valley pseudospin and drives domain wall into motion.
We describe such wall dynamics by studying linear response around the fixed points: $\delta \vec{\chi}\equiv =\vec{\chi}-\vec{\chi}^*$. As discussed in the maintext, the edge state relaxation is concentrated at the hotspot which we locate at $y=0$. Hence, we seek a solution of the form $\delta \vec{\chi} \propto e^{-qy}$ where $q=0$ correspond to the thermodynamic result discussed in the maintext. Linearizing Eq.~\eqref{eq:SX_n_phi} and \eqref{eq:SX_n_phi_2} around the fixed points yield:
\begin{equation}
    \hbar\frac{d}{dt}\begin{pmatrix} \delta \phi \\
\delta X /\lambda
\end{pmatrix}=\frac{A_M}{1+\alpha_{X}\alpha_{\phi}}
\begin{pmatrix}2 \alpha_{\phi} (\pi A q^2 + (-1)^n K_\perp A_{M}^{-1})  & 4Aq^{2}-k\lambda\\
-2(\pi A q^2 + (-1)^n K_\perp A_{M}^{-1}) & \alpha_X (4Aq^{2}-k\lambda) 
\end{pmatrix}\begin{pmatrix} \delta \phi \\
\delta X /\lambda 
\end{pmatrix} \equiv  R\cdot \delta \vec{\chi}.
\end{equation}
We tabulate the stability analysis around the fixed points in Table.~1. The rate matrix $R$ has the following eigenvalues and eigenvectors: 
\begin{equation}
    \epsilon_{\pm} = \frac{a \alpha_{\phi}  + b \alpha_{X} \pm \sqrt{a^2 \alpha_{\phi}^2  + b^2 \alpha_{X}^2 - 2 ab(2+\alpha_{\phi}\alpha_{X}) }  }{2(1+\alpha_{\phi}\alpha_{X})} \;\; ,\; \;    
    \vec{u}_{\pm} = (u_{\pm},v_{\pm})^{\text{T}}= \left( \frac{1+\alpha_{\phi}\alpha_{X} }{a}  \epsilon_{\pm} -\alpha_{\phi} \,,\, 1 \right)^\text{T},
\end{equation}
where $a=2A_M(\pi A q^2 + (-1)^n K_\perp A_{M}^{-1})$ and $b=4A_MAq^{2}-kA_M\lambda$. Next, we construct a traveling domain wall from the superposition of $\vec{u}_1$ and $\vec{u}_2$:
\begin{equation}
    \vec{\chi}(y,t) = \vec{\chi}^* + \sum_{i=\pm} c_i\, \vec{u}_i e^{\epsilon_it/\hbar-qy},
\end{equation}
where the coefficients $c_{\pm}$ are determined from the initial conditions located at the hotspot:
\begin{equation} \label{eq:BC}
    \partial_t \vec{\chi}(0,0) = \left(\delta\mu/\hbar \,, \, v_0/\lambda \right)^{\text{T}}.
\end{equation}
Here $\delta \mu = \gamma eV$ where $\gamma <1$ is the hotspot equilibration parameter and $eV$ is the voltage difference between the two contacts that sourced valley edge states.
The initial velocity $v_0=\dot{X}(0,0)$ is a phenomenological parameter that depends on the microscopic details of the hotspot. From the initial conditions, we find
\begin{equation}
    c_{\pm} = \frac{ \delta \mu -\hbar v_0\lambda^{-1}\, u_{\mp}}{u_{\pm}-u_{\mp}},
\end{equation}
and the wall position is given by 
\begin{equation} \label{eq:X_fin}
    X(y,t)=X^*+ \frac{\lambda e^{-qy} }{u_+ - u_- } 
    \bigg[ 
    \delta \mu \, (e^{\epsilon_+ t /\hbar} -e^{\epsilon_- t /\hbar})
    -\hbar v_0 \lambda^{-1} (u_+ e^{\epsilon_+ t /\hbar} + u_- e^{\epsilon_- t /\hbar}) \bigg].
\end{equation}
In Fig.~2 of the maintext, we plot the dynamics of vertical wall-profile from Eq.~\eqref{eq:X_fin} in the small time limit: $\delta X(y,t) =   \tilde{v}\,t e^{-qy}$ where $\tilde{v}$ is an effective wall velocity.

\begin{table}[h]
\label{table}
\centering
\begin{tabular}{|c|c|c|}
\hline 
 & Odd $n$ & Even $n$\tabularnewline
\hline 
\hline 
Strong Pinning $(k>4Aq^{2}/\lambda)$ & $\begin{cases}
\text{stable} & K_{\perp}>\pi Aq^{2}\\
\text{unstable} & K_{\perp}<\pi Aq^{2}
\end{cases}$ & unstable\tabularnewline
\hline 
Weak Pinning $(k<4Aq^{2}/\lambda)$ & unstable & unstable\tabularnewline
\hline 
\end{tabular}
\caption{Linear stability analysis. For even $n$, the dynamical system is generically unstable. When the pinning is weak $4Aq^2>k\lambda$, the fixed points are unstable spirals or unstable nodes and when the pinning is strong $4Aq^2<k\lambda$, the fixed points are saddle-points. For odd $n$ fixed points with weak perpendicular anisotropy $\pi A q^2 < K_\perp$, the dynamical stability is the same as even $n$ fixed points. However, for odd $n$ fixed points with strong perpendicular anisotropy $\pi A q^2 > K_\perp$, the domain wall can either be a saddle point if pinning is weak  $4Aq^2>k\lambda$,  or a stable fixed point if the pinning is strong $4Aq^2<k\lambda$. }
\end{table}

\end{document}